\def\sqiggt{\hbox{\rlap{\lower.55ex \hbox {$\sim$}}
\kern-.3em \raise.4ex \hbox{$>$}\,}}
\def\sqig{$\sim\,$}  \def\mdot{$\dot{M}$}
\documentstyle{mn}
\title[Cataclysmic variables]
{On the orbital period distribution of cataclysmic variables}
\author[C.~Hellier \&\ T.~Naylor]{Coel Hellier and T.~Naylor \\
Department of Physics, Keele University, Keele, Staffordshire, ST5 5BG}

\date{ }
\begin{document}
\maketitle
\begin{abstract}
Using the latest compilation of cataclysmic variable orbital periods
by Ritter \&\ Kolb we argue against Verbunt's conclusion that 
the period gap is not significant for nova-like variables.
We also discuss the relation of the VY~Scl stars to the dwarf novae.
\end{abstract}

 
\section{Introduction}
Two fundamental characteristics of a cataclysmic variable (CV) are the
orbital period, $P$, typically between 80 m and 9 h, and the behaviour of
the optical lightcurve, showing either the recurrent 2--5 mag outbursts of a
dwarf-nova (DN), or the steady lightcurve of a nova-like variable (NL). DN
are thought to be low accretion rate (\mdot) systems with accretion discs
cool enough to undergo hydrogen ionization instabilities, whereas NLs are
always too hot for this to occur (see Warner 1995 for a comprehensive
review of the whole field). The NLs predominantly have $P$\,$>$\,3 h,
implying a high \mdot\ and thus high angular momentum loss from the binary
in this range. Systems with $P$\,$<$\,2 h are predominantly DN, implying a
lower angular momentum loss from gravitational radiation alone. Fewer
systems occur in the `period gap' between 2 and 3 h, a possible consequence
of the mechanism for additional braking switching off at \sqig 3 h. 

However, Verbunt (1997) has concluded that the period gap is not significant
for NLs, and thus queried whether the additional braking mechanism (usually
suggested to be braking by the magnetic field of the secondary) is required.
In this paper we re-examine the issue using Ritter \&\ Kolb's (1998) more
recent list of CVs with known orbital period. Further, we address
the nature of VY~Scl stars, which are currently poorly understood
(e.g.~Livio \&\ Pringle 1994; Wu, Wickramasinghe \&\ Warner 1995; Verbunt
1997).

\section{The sample}
The philosophy of this paper is that we can use the outburst properties
recorded by Ritter \&\ Kolb (1998) as an indicator of \mdot\
(c.f.~Shafter 1992). Thus we created a sample of high \mdot\ systems into
which we placed the NLs. Then, in a divergence from the method of Verbunt
(1997), we removed all the NLs listed by Ritter \&\ Kolb (1998) as certain
AM~Her stars, DQ~Her stars, or intermediate polars.  This is because AM~Hers
don't possess discs and thus the presence  or absence of outbursts is no
longer an indicator of \mdot\ (the use of the term `nova-like' for AM Her
stars is more historical than physical). Similarly, while intermediate
polars mostly have discs, the magnetic field also affects the outburst
behaviour (e.g.\ Warner 1996; Hellier, Mukai \&\ Beardmore 1997). We then
created versions of the `high \mdot' sample both including and excluding old
novae, in case the nova eruption leaves the CV with an \mdot\ temporarily
atypical of its $P$. Lastly, since we wanted to investigate the status of
VY~Scl stars we removed them from the NL sample and placed them into a
separate `VY' sample. We refer to the high \mdot\ sample as `nNL' for
`normal NLs' (i.e. no magnetics or VY~Scl stars) or as `nNL+N' when it
includes the old novae (recurrent novae were excluded from both samples).

A further reason for omitting old novae and magnetic systems is the
potential bias due to selection effects: nova explosions draw attention to
the underlying CV, and the majority of magnetic systems are first seen in
X-rays, whereas non-magnetic CVs are nearly all discovered optically (the
variability of DN also make them more obvious than NLs, but as long as this
affects all orbital periods equally it won't affect our analysis).

The Z~Cam stars are hybrids showing periods of DN outbursts and periods
of `standstill' in which they act as NLs. The accepted explanation 
(e.g.~Warner 1995) is that their \mdot\ is finely poised at the
boundary between NL and DN behaviour, so that a minor excursion moves
them from one class to the other. We place these stars in a medium \mdot\
sample called `ZC'.
Our low-\mdot\ sample, essentially the DN, contains all systems that can
never sustain an excursion to the hot side of the disc instability. We call
these `nDN' for `normal DN', to denote our exclusion of the Z~Cams.

\section{Is there a period gap?}
In Fig.~1 we plot the cumulative period distribution for our low-\mdot\
sample (nDN) against the sample of all medium and high \mdot\ systems (i.e.
nNL+ZC+VY). There are clear breaks at 2.1 h (in nDN) and \sqig 3 h (in 
nNL+ZC+VY) and a deficit of systems 
in-between.

\begin{figure}\vspace{8.2cm}   % Fig 1
\caption{The cumulative orbital period distribution of low \mdot\
CVs (nDN) compared with that of medium and high \mdot\ CVs (nNL+ZC+VY).
The dashed line is the nNL+N sample.}
\includegraphics{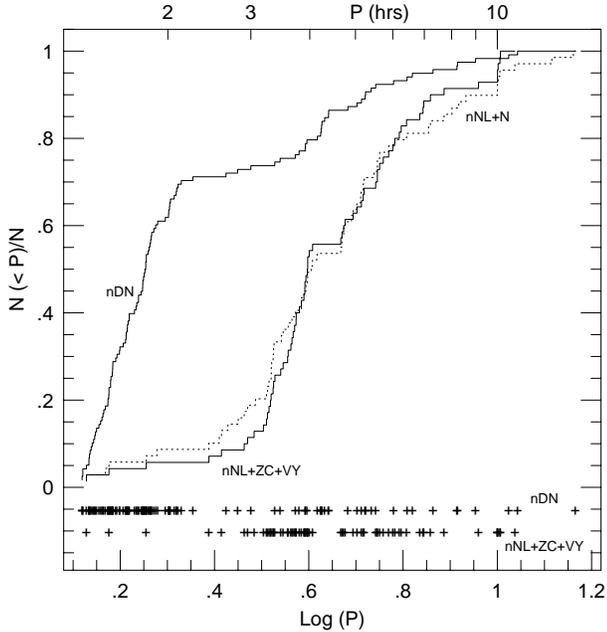}
\end{figure}

In case the reader is suspicious of our concocted high+medium-\mdot\
sample, we first test the significance of the upper edge of the period
gap with the purer nNL sample. A one-sided Kolmogorov-Smirnov (K-S)
test shows that the distribution of nNL between 6 and 3 h is compatible
with arising from a parent distribution of constant probability in
$\log P$ at 67
per cent confidence. When testing against a constant distribution from 6
h to 2.1 h (where the DN turn on) the probability drops to 2 per cent,
and when testing down to the 80-m period minimum the probability drops
to $10^{-4}$.  For the other samples the drops in probability against a
constant distribution in $\log P$ when the range is extended from 7--3 h to 7--2.1 h
are: nNL+N, 53 to 1 per cent; nDN, 43 to 2 per cent; and for nDN+ZC, 41
to 0.2 per cent. Thus all samples, including samples which are
independent, show a significant ($>$95 per cent) reduction in the number
of CVs below \sqig 3 h. This  conclusion differs from that of Verbunt
(1997) because he included magnetic systems amongst the NLs. The
evolution of magnetic systems may well be different from that of
non-magnetics (e.g.~Wickramasinghe \&\ Wu 1994), and using a two-sided
K-S test for the current sample of systems with $P$\,$<$\,5 h gives only
a 15 per cent probability that the distribution of AM~Her stars and that
of all non-magnetic CVs arise from the same parent  population.

Turning now to shorter periods, the NLs are too few to investigate the
significance of the break at 2 h, so we do this with the low and medium
\mdot\ systems. Both nDN and nDN+ZC are consistent with a constant
distribution in $\log P$ between 80 m and 2.1 h at 32 per cent probability,
but have only a 10$^{-6}$ probability of constant distribution up to 3 h.
Thus the numbers of DN decline at 2 h, significantly before the rise in NL
numbers at 3 h.

Verbunt (1997) has noted that the gap is most significant when using
different samples above and below the gap (NL and DN respectively),
raising the possibility of awkward selection effects. However, both the
nDN and nDN+ZC samples, which are homogeneous observational samples of
all systems reliably showing DN outbursts, show both the 2 h break and
the 3 h break by themselves ($>$98 per cent confidence).
Thus, regardless of the merits of the magnetic braking theory, the
period gap is an observational fact. Since the orbital cycle is not used
in the initial discovery of most CVs, selection effects could only
operate if periods of certain lengths were much harder to detect in known
CVs. While lengths comparable to or longer than a night's observing
($P$\sqiggt 6 h) may indeed be harder to detect, we can
think of no reason why periods of 2.5 h would be less obvious than those
of 1.5 h or 3.5 h.

\section{The nature of VY~Scl stars}
Considering only the stars above the gap ($P$\,$>$\,3 h), 
the samples nNL, nDN and ZC are all compatible with arising from the same
parent distribution (at 55 per cent probability for nNL and nDN, and at 
67 and 87 per cent probabilities for ZCs against nNLs and nDN
respectively).  VYs
occur preferentially just above the period gap, with all but two systems
(whose periods are uncertain) in the range 3\,$<$\,$P$\,$<$\,4 h.

Testing the low \mdot\ sample (nDN) against higher \mdot\ systems
(nNL+ZC+VY), provides evidence for a deficit of DN in the 3--4 h  range
(Fig.~2). Shafter (1992) reported this at 98 per cent 
significance. However, because of the new discoveries included in our
sample, a two-sided K-S test now gives a 17 per cent probability that
they come from the same parent distribution. Hence, while we need an
explanation for the occurrence of VYs in this range, the lack of DN 
might be real, or might simply be a chance occurrence.

\begin{figure}\vspace{8.2cm}   % Fig 2
\caption{The distribution of low \mdot\ CVs (nDN) above the 
gap, compared with that of higher \mdot\ CVs (nNL+ZC+VY).}
\includegraphics{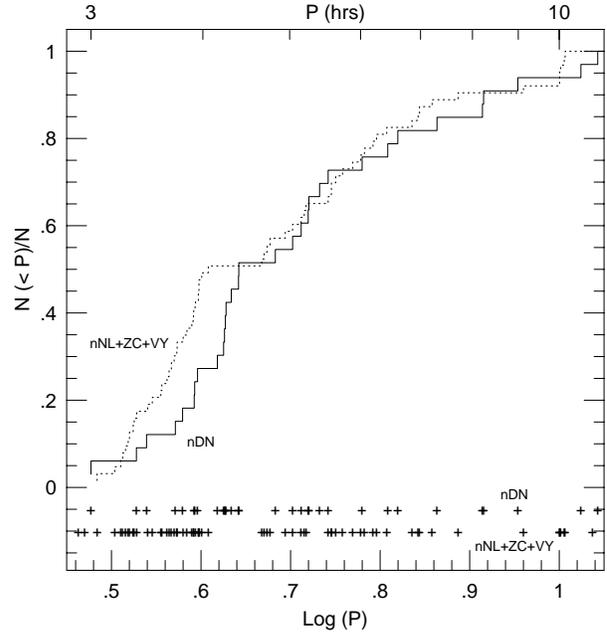}
\end{figure}

\begin{figure*}\vspace{5.7cm}   % Fig 3
\caption{Longterm lightcurves of Z~Cam and TT~Ari, a typical
VY~Scl star. The relatively high \mdot\ in Z~Cam ensures frequent, regular
outbursts when not in standstill. In contrast, the low-state of TT~Ari 
is much lower and longer lasting, with few or no DN eruptions (carets
denote upper limits). The data are
1-day averages based on compilations by the AAVSO, VSOLJ \&\ AFOEV.}
\includegraphics{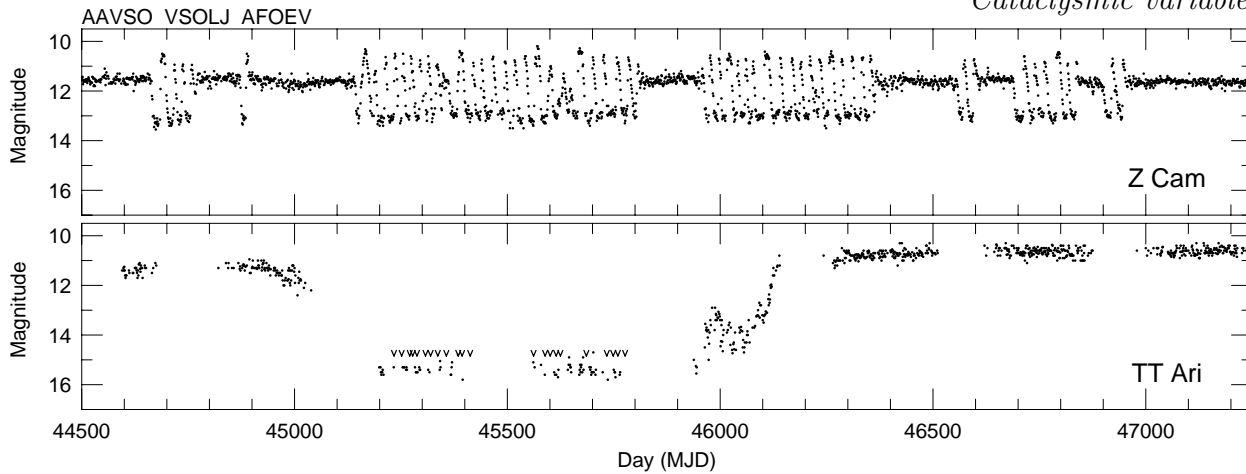}
\end{figure*}

One suggested explanation is that DN turn into VYs in the 3--4 h range
(see Livio \&\ Pringle 1993 and Verbunt 1997). We now argue that this is
unlikely. The VYs have long stretches acting as NLs, and have absolute
magnitudes typical of NLs rather than DN (e.g.~Warner 1995). Thus to
turn a DN into a VY we need to increase \mdot. However, we think we know
what happens to a DN as \mdot\ increases: it turns into a ZC, and then
into a NL. Thus VYs would have to be an intermediate stage similar to
ZCs, perhaps between ZC and NL. However, the transition DN to ZC is marked
by a trend to lower-amplitude, more frequent and more regular
outbursts, with less time spent at minimum.  
The transition ZC to NL then sees periods of standstill
increasing in frequency and length until they become all we see. 

As remarked by Warner (1995), the VYs are distinctly different
(see Fig.~3). The lightcurve changes have a greater amplitude (with an
average $\Delta$mag of 4.5 compared with 2.5 for ZCs; data from Ritter \&\
Kolb 1998); the low states are erratic and unpredictable; the star can spend
months in a low state; such low states can have an \mdot\ well below that of
a ZC at minimum; the transitions between high and low states often take far
longer than DN transitions, lasting up to \sqig 1 yr; and, in contrast to
ZCs, outbursts are rare or absent when in a low state.  Further, if VYs are
objects transitional in \mdot, similar to ZCs, why don't they occur at all
periods above the gap, as NLs, ZCs and DN do? Why do VYs increase at
3\,$<$\,$P$\,$<$\,4 h whereas ZCs don't? And why is there no transitional
object between a ZC and a VY? 

From the dissimilarity of VY variability to DN and ZC
variability, we conclude that we need a mechanism other than the disc
instability to explain VYs. A big clue is that AM~Her stars show low
states very similar to those in VYs, and since they
don't have discs the low states must involve changes in the mass 
transfered from the secondary star. Possibilities include the star-spot
mechanism of Livio \&\ Pringle (1994), and irradiation-driven
mass transfer cycles (Wu, Wickramasinghe \&\ Warner 1995). Irradiational
heating of the secondary can increase the \mdot\ above the rate justified by
angular momentum loss, with the \mdot\ leading to enhanced irradiation. This
unstable feedback must break, plunging the system into a low state of near
zero \mdot\ as the secondary cools (see also Warner 1995 and Hellier 1996
for applications of this idea to VYs and SW~Sex stars). The mechanism would
require a high secular \mdot, and is more efficient in a close binary,
producing the restriction of VYs to the period range just above the gap. In
AM~Hers the secondary is not shielded by the disc, allowing irradiation to
be effective at lower \mdot s.

If the above is correct, the deficit of DN with 3\,$<$\,$P$\,$<$\,4 h 
would have to be explained by a different mechanism, or dismissed as chance.
As discussed by Shafter (1992), a general increase in \mdot\ could remove
DN, but would have to be reconciled with theories of the width of the period
gap.

\section{Conclusions}
This paper has been largely a defence of orthodoxy, showing
that the period gap in cataclysmic variables is significant 
and that the nova-like variables show a cutoff at \sqig 3 hr.
Further, we've argued against Verbunt's (1997)
classification of VY~Scl stars with the dwarf novae. Verbunt suggested 
that since VY~Scl variability is probably caused by changes in mass
transfer from the secondary star, this could also have a major role
in DN outbursts. We've shown that the differences are sufficient to
require separate mechanisms, such as the disc instability for DN and
irradiation-driven mass transfer cycles for VY~Scl stars.

\section*{Acknowledgments}
We thank Frank Verbunt for helpful comments on this work and 
gratefully acknowledge the contribution of amateur light-curve estimates
complied by the AAVSO, VSOLJ and AFOEV. TN is a PPARC Advanced Fellow.
%American Association of Variable Star Observers,
%the Variable Star Observers League of Japan, and the Association
%Francaise des Observateurs d'Etoiles Variables.


\begin{thebibliography}{}
\bibitem[]{}Hellier C., 1996, ApJ, 471, 949
\bibitem[]{}Hellier C., Mukai K., Beardmore A.\,P., 1997, MNRAS, 292, 397
\bibitem[]{}Livio M., Pringle J.\,E., 1994, ApJ, 427, 956
\bibitem[]{}Ritter H., Kolb U., 1998, A\&AS, in press
\bibitem[]{}Shafter A.\,W., 1992, ApJ, 394, 268
\bibitem[]{}Verbunt F., 1997, MNRAS, 290, L55
\bibitem[]{}Warner B., 1995, Cataclysmic variable stars, Cambridge 
    University Press, Cambridge
\bibitem[]{}Warner B., 1996, Ap\&SS, 241, 263
\bibitem[]{}Wickramasinghe D.\,T., Wu K., 1994, Ap\&SS, 211, 6
\bibitem[]{}Wu K., Wickramasinghe D.\,T., Warner B., 1995,
Publ. Astron. Soc. Aust., 12, 60
\end{thebibliography}
\end{document}